\documentstyle[manuscript,amssymb,aps,prb]{revtex}
\begin{document}
\title{
Common vacuum conservation amplitude in the theory of the
radiation of mirrors in two-dimensional space-time and of charges
in four-dimensional space-time
}
\author{V. I. Ritus\cite{*)}}
\address{P. N. Lebedev Physics Institute, Russian Academy of
Sciences, 117924 Moscow, Russia}
\maketitle

\begin{abstract}
The action changes (and thus the vacuum conservation
amplitudes) in the proper-time representation are found for an
accelerated mirror interacting with scalar and spinor vacuum
fields in $1+1$ space.  They are shown to coincide to within the
multiplier $e^2$ with the action changes of electric and scalar
charges accelerated in $3+1$ space.  This coincidence is
attributed to the fact that the Bose and Fermi pairs emitted by a
mirror have the same spins 1 and 0 as do the photons and scalar
quanta emitted by charges.  It is shown that the propagation of
virtual pairs in $1+1$ space can be described by the causal
Green's function $\Delta_f(z,\mu)$ of the wave equation for $3+1$
space.  This is because the pairs can have any positive mass and
their propagation function is represented by an integral of the
causal propagation function of a massive particle in $1+1$  space
over mass which coincides with $\Delta_f(z,\mu)$.  In this integral
the lower limit $\mu$ is chosen small, but nonzero, to eliminate
the infrared divergence.  It is shown that the real and imaginary
parts of the action change are related by dispersion relations,
in which a mass parameter serves as the dispersion variable.
They are a consequence of the same relations for $\Delta_f(z,\mu)$.
Therefore, the appearance of the real part of the action change
is a direct consequence of the causality, according to which
$\mathop{\rm Re}\Delta_f(z,\mu)\neq 0$ only for timelike and zero
intervals.
\end{abstract}

\renewcommand{\thesection}{\arabic{section}}
\section{INTRODUCTION}

An intriguing symmetry between the creation of particle
pairs by an accelerated mirror in $1+1$ space and the emission of
single quanta by a charge accelerated as a mirror in $3+1$ space
was discovered in Refs. \onlinecite{1,2,3}.  This symmetry is
confined to coincidence of the spectra of the Bose and Fermi
pairs created by the mirror with the spectra of the photons and
scalar quanta emitted by electric and scalar charges, if the
doubled frequencies $\omega$ and $\omega '$ of the quanta in a pair
created
by the mirror are identified with the components $k_\pm=k^0\pm
k^1$ of the 4-wave vector $k^\alpha$ of the quantum emitted by the
charge:
\begin{equation}
2\omega=k_+,\quad 2\omega '=k_-.
\eqnum{1}
\end{equation}
It was shown in Ref. \onlinecite{3} that the Bogolyubov
coefficients $\beta^B_{\omega '\omega}$ and
$\beta^F_{\omega '\omega}$, which
describe the spectra of Bose and Fermi radiation of a mirror,
 are related to
the Fourier transforms of the 4-current density $j_\alpha(k_+,k_-)$
and the scalar charge density $\rho(k_+,k_-)$, which describe the
spectra of photons and scalar quanta emitted by charges, by the
following expressions\cite{1)}
\begin{equation}
\beta^{B*}_{\omega '\omega}=-\sqrt{\frac{k_+}{k_-}}\frac{j_-}{e}=
\sqrt{\frac{k_-}{k_+}}\frac{j_+}{e},  \eqnum{2}
\end{equation}
\begin{equation}
\beta^{F*}_{\omega '\omega}=\frac{1}{e}\rho(k_+,k_-).  \eqnum{3}
\end{equation}
It was also shown that $\beta^*_{\omega '\omega}$ is the source
amplitude of a
pair of particles which are only potentially emitted to the right
and to the left with the frequencies $\omega$ and $\omega '$.
In other
words, it is the virtual-pair creation amplitude.  The pair
becomes real when one of its particles undergoes internal
reflection with a frequency change and both particles move in the
same direction, i.e., to the right in the case of a right-sided
mirror and to the left in the case of a left-sided mirror.
Therefore, for a right-sided mirror, for example, the emission
amplitude $\langle out\,\omega '' \omega|in \rangle$ of a real pair
of
particles with the frequencies $\omega$ and $\omega ''$ is
connected with
the virtual-pair creation amplitude $\beta^*_{\omega '\omega}$
by the relation
\begin{equation}
\langle out\,\omega ''\omega|in \rangle=-\sum\limits_{\omega '}
\langle out\,\omega ''|\omega '\,in
\rangle \beta^*_{\omega '\omega},\eqnum{4}
\end{equation}
where
$\langle out\,\omega ''|\omega '\,in \rangle$ is the amplitude of
single-particle scattering on the mirror.  The energy and
momentum of this
real pair equal $\omega+\omega ''$ and $\omega+\omega ''$,
i.e., the pair does not have mass, nor do its components.

A virtual pair is another matter.  According to (1), the
zeroth and first components of the 4-momentum $k^\alpha$ of a quantum
emitted by a charge are equal to the energy and momentum of a
virtual pair of massless particles created by a mirror:
\begin{equation}
k^0=\omega+\omega ',\quad
k^1=\omega-\omega ', \eqnum{5}
\end{equation}
and form the timelike 2-momentum of the pair in $1+1$ space.
Clearly, the quantity
\begin{equation}
m=\sqrt{k_+k_-}=2\sqrt{\omega\omega'},    \eqnum{6}
\end{equation}
being an invariant of Lorentz transformations along axis 1, is
the mass of the virtual pair, and, at the same time, it equals
the transverse momentum $k_\perp=\sqrt{k^2_2+k^2_3}$ of the
massless real quantum emitted by the charge.

The fact
that the source amplitude $\beta^B_{\omega '\omega}$ of a virtual
pair of bosons is specified by the current $j^\alpha(k_+,k_-)$, while
the source
amplitude $\beta^F_{\omega '\omega}$ of a virtual pair of fermions is
specified by the scalar $\rho(k_+,k_-)$, means that the spin of a
boson pair equals 1, while the spin of a fermion pair equals
zero.  Thus, the coincidence between the emission spectra of a
mirror in $1+1$ space and charges in $3+1$ space can be
attributed to the coincidence between the moment of a pair
emitted by the mirror and the spin of a particle emitted by the
charge.\cite{3}

The relation (2) can be written in the explicitly invariant
form
\begin{equation}
e\beta^{B*}_{\omega '\omega}=\varepsilon_{\alpha\beta}k^\alpha
j^\beta/\sqrt{k_+k_-},    \eqnum{7}
\end{equation}
and, more specifically, in the form of the scalar product of the
2-current
vector $j^\beta$ and the 2-polarization pseudovector $a_\beta$
of a Bose pair
\begin{equation}
a_\beta=\frac{\varepsilon_{\alpha\beta}k^\alpha}{\sqrt{k_+k_-}},\quad
a_0=-\frac{k^1}{\sqrt{k_+k_-}},\quad
a_1=\frac{k^0}{\sqrt{k_+k_-}}.  \eqnum{8}
\end{equation}
The spacelike pseudovector $a_\beta$ is constructed from the zeroth
and first components of the 4-momentum $k^\alpha$ of the quantum
emitted by the mirror.  It is orthogonal to the 2-momentum of the
pair, has a length equal to 1, and is represented in the comoving
frame of the pair only by a spatial component, as is the current
vector $j^\alpha$.

In this work we find the vacuum conservation amplitude for
acceleration of a mirror, which is defined by the change $\Delta W$ in
the self-interaction of the mirror due to its acceleration.  The
problem here is essentially finding $\mathop{\rm Re}\Delta W$ from the
previously found quantity $\mathop{\rm Im}\Delta W$, whose doubled
value coincides in a certain approximation (see below) with the
mean number of real pairs formed by the mirror.  Three different
methods are used for this purpose.

The first (and principal) method is considered in Sec. 2 and
consists in transforming the original space-time representation for
the mean number of pairs into a proper-time representation,
whose kernel turns out to be the relativistically invariant
singular even solution $(1/2)D^1(z)$ of the wave equation in
$3+1$ space.  Then, the function $D^1(z)$ in the expression
obtained for the number of pairs is replaced by the even solution
$\Delta^1(z,\mu)$
of the Klein--Gordon equation in order to invariantly
and symmetrically eliminate the infrared divergence in the
integral for the number of pairs using the small mass parameter
$\mu$ instead of the large trajectory-length parameter $L$ used
in the original expression.  The parameters $\mu,
L^{-1}\ll\kappa$, if $\kappa$ is the characteristic acceleration
on the trajectory.  Finally, treating the function
$(1/2)\Delta^1(z,\mu)$
as the imaginary part of the kernel defining $\Delta
W$,
by analytic continuation with respect to $z^2$ we can reconstruct a
relativistically
invariant and even in $z$ kernel which coincides with the
causal Green's
function $\Delta_f(z,\mu)$ specific to $3+1$ space. The resultant
action changes of
a mirror and a charge differ only by the multiplier $e^2$, and
the interactions
are described by the same causal propagation function.  Thus,
the difference in dimensionality of the spaces is compensated by the
difference in the mechanism of interaction transfer: it is realized by
pairs in $1+1$ space and by individual particles in $3+1$ space.

Section
3 presents a direct calculation of the self-interaction changes
$\Delta W_f^{B,F}$ for a concrete, but fairly
general mirror trajectory.  The invariant functions of the
relative velocity of the trajectory ends obtained for $\Delta
W_f^{B,F}$ are consistent with the results of Sec. 2.

In Sec. 4  $\mathop{\rm Re}\Delta W_f$ is reconstructed from
$\mathop{\rm Im}\Delta W_f$ using dispersion relations, in which $\mu$
appears as the dispersion variable.  It is shown that the
dispersion relations for $\Delta W_f$ are a consequence of the same
relations for $\Delta_f(z,\mu)$ with timelike $z$ as the parameter.
As a consequence of causality only for such $z$ the values of
$\mathop{\rm Re}\Delta_f$ and $\mathop{\rm Re}\Delta W_f$
are nonzero and
are connected with $\mathop{\rm Im}\Delta_f$ and $\mathop{\rm
Im}\Delta W_f$, respectively, by the dispersion relations.

The fifth section examines other analytic continuations of
$i\Delta^1/2$
onto the real $z^2$ axis that lead to kernels for $\Delta W$
whose real parts are not even in $z$.

A physical interpretation of the results is presented in the
sixth, concluding section.  The appearance of a causal
propagation function characteristic for four-dimensional
space-time in two-dimensional space-time is attributed to
interaction transfer by pairs of different mass.

\section{PROPER-TIME REPRESENTATION OF THE CHANGE OF ACTION}

The following representations were obtained in Ref.
\onlinecite{2} for mean numbers of radiated Bose and Fermi
particles:
\begin{equation}
N^{B,F}=\frac{1}{4\pi^2}\int\limits_{-\infty}^{\infty}
du\,K^{B,F}(u), \eqnum{9}
\end{equation}
\begin{equation}
K^B(u)=\mathop{\diagup\!\!\llap{$\displaystyle\int$}}
\limits_{-\infty}^{\infty}
\frac{dv}{v-f(u)}
\left[\frac{1}{g(v)-u}-\frac{f'(u)}{v-f(u)}  \right],
\eqnum{10}
\end{equation}
\begin{equation}
K^F(u)=-\sqrt{f'(u)}
\int\limits_{-\infty}^{\infty}
\frac{dv}{v-f(u)}
\left[\frac{\sqrt{g'(v)}}{g(v)-u}-\frac{\sqrt{f'(u)}}{v-f(u)}
\right].\eqnum{11}
\end{equation}
It follows from these representations  for trajectories with the
asymptotically constant velocities $\beta_1$ and $\beta_2$ at the ends
and a nonzero Lorentz-invariant relative velocity
\begin{equation}
\beta_{21}=\frac{\beta_2-\beta_1}{1-\beta_2\beta_1},\quad
\theta= \tanh^{-1} \beta_{21},     \eqnum{12}
\end{equation}
that the mean number of massless quanta emitted is infinite
(there is infrared divergence).  In fact, in this case it follows
from formulas (10) and (11) for $u \rightarrow \pm\infty$ (more
precisely, for $|u|\gg\kappa^{-1}$, i.e., outside the region
where the mirror experiences the characteristic acceleration
$\kappa$) that the functions $K^{B,F}(u)$ possess universal
behavior, which depends only on $\beta_{21}$:
\begin{equation}
K^B(u)\approx
\pm\frac{1}{u}\left( \frac{\theta e^{\mp \theta}}{\sinh \theta}-1
\right)=\pm\frac{1}{u}
\left( \frac{\theta}{\tanh \theta}
-1 \right)-\frac{\theta}{u}, \eqnum{13}
\end{equation}
\begin{equation}
K^F(u)\approx \pm\frac{1}{u}\left( 1-\frac{\theta}{\sinh \theta}
\right).\eqnum{14}
\end{equation}
The relativistically invariant coefficients accompanying $u^{-1}$ are
formed on the parts of the trajectories with
asymptotically
constant velocities.  As a result, the mean number of quanta
emitted on the part of the trajectory covering the acceleration
region grows logarithmically as the length $2L$ of that part
is increased:
\begin{equation}
N^B=\frac{1}{2\pi^2}\left(
\frac{\theta}{\tanh \theta}-1 \right)\ln(L\kappa)+
2b^B(\theta),   \eqnum{15}
\end{equation}
\begin{equation}
N^F=
\frac{1}{2\pi^2}\left(
1-\frac{\theta}{\sinh \theta} \right)\ln(L\kappa)+
2b^F(\theta),\quad
L\kappa\gg 1.       \eqnum{16}
\end{equation}
Let us focus our attention on the fact that the odd (with respect
to both $u$ and $\theta$)
term in the asymptotics of $K^B(u)$ does not make
a contribution to the integral defining $N^B$.  The terms
$2b^{B,F}$ do not depend on $L$ if $L\kappa\gg 1$, but they can
depend on the specific form of the trajectories.

We note that there are representations for $N^{B,F}$ which
differ from (9)--(11) by mirror symmetry, i.e., by the
replacements $u \rightleftarrows v$ and $f(u) \rightleftarrows
g(v)$.  The integrands $K^{B,F}(v)$ defining them differ from
$K^{B,F}(u)$, but are denoted below by the same letter, since
they
are values of the same functional taken for two mirror-symmetrical
trajectories:  $K(u)\equiv K[u;g]$ and $K(v)\equiv K[v;f]$.  As
$v \rightarrow \pm\infty$, $K^{B,F}(v)$ have asymptoticses which
differ from (13) and (14) by the replacements $u \rightarrow v$
and $\theta \rightarrow -\theta$.

The vacuum conservation amplitude of an accelerated mirror
is specified by the action change $\Delta W=W|^F_0$ (i.e., the
difference between the actions for the accelerated and
unaccelerated mirror) and has the form $\exp(i\Delta W)$, where
$2\mathop{\rm Im}\Delta W=N$, if the interference effects in the
creation of two or more pairs are neglected.  We shall consider
particle and antiparticle to be nonidentical; otherwise, in
the same approximation $2\mathop{\rm Im}\Delta W=(1/2)N$ (see Ref.
\onlinecite{3}).

Now the main task is to find $\mathop{\rm Re}\Delta W$.  For this
purpose, we obtain a suitable representation for $\mathop{\rm
Im}\Delta W$ and utilize relativistic-invariance and causality
arguments.

Let us consider the space-time representation for $N$ which
was the direct ``parent'' of the representation (9)--(11) [see Ref.
\onlinecite{2}].  In this representation
\begin{equation}
N^B=\mathop{\int\!\!\!\int} \limits_{-\infty\quad}^{\quad\infty}
du\,dv\,S(u,v)\big|^F_0,\quad
S(u,v)=\frac{1}{8\pi^2}\left[
\frac{1}{\left( v{-}f(u){-}i\varepsilon \right)\left( g(v)-u{-}i
\delta \right)} + {\rm c.c.}
  \right].      \eqnum{17}
\end{equation}
We go over from the independent characteristic variables $u$ and
$v$ to the proper-time moments $\tau$ and $\tau '$ of two
points on the world trajectory of the mirror $x^\alpha(\tau)$:
\begin{equation}
u=x^0(\tau)-x^1(\tau)=x_-(\tau),\quad
v=x^0(\tau ')+x^1(\tau ')=x_+(\tau '). \eqnum{18}
\end{equation}
Then
\begin{equation}
f(u)=x^0(\tau)+x^1(\tau)=x_+(\tau),\quad
g(v)=x^0(\tau ')-x^1(\tau ')=x_-(\tau '),    \eqnum{19}
\end{equation}
and $S(u,v)$ becomes a relativistically invariant function of the
two-dimensional
vector $z^\alpha=x^\alpha(\tau)-x^\alpha(\tau ')\equiv (x-x')^\alpha$
joining the points
$x^\alpha=x^\alpha(\tau)$ and $x'{}^\alpha=x^\alpha(\tau ')$
on the mirror trajectory:
\begin{eqnarray}
S(z)&=&
\frac{1}{8\pi^2}\left[
\frac{1}{(x'_+-x_+-i\varepsilon)(x'_--x_--i\delta)}+{\rm c.c.}
  \right]=
                \nonumber\\
&=&
\frac{1}{8\pi^2}\left[
\frac{1}{z_+z_-+i\varepsilon\mathop{\rm sgn} z^0}+{\rm c.c.}
  \right]=
\frac{1}{8\pi^2}\left[
\frac{1}{-z^2+i\varepsilon\mathop{\rm sgn} z^0}+{\rm c.c.}
  \right]=
-P\frac{1}{4\pi^2z^2}.
                        \eqnum{20}
\end{eqnarray}
The individual terms in (20) and their sum are well-known
relativistically invariant singular functions in quantum
electrodynamics (we use the notation in Ref. \onlinecite{4}, but
our $D^1$ and $\Delta^1$ lack the multiplier $i$):
\[
D^\pm(z)=\frac{\pm i}{4\pi^2(z^2\pm i\varepsilon\mathop{\rm sgn}
z^0)}=
\frac{1}{4\pi^2}\left[\pi\varepsilon(z^0)\delta(z^2)\pm\frac{i}{z^2}
\right],
\]
\begin{equation}
\eqnum{21}
\end{equation}
\[
D^1(z)=\frac{1}{2\pi^2z^2},
\]
so that
\begin{equation}
S(z)=-\frac{i}{2}\left[D^-(z)-D^+(z)  \right]=-\frac{1}{2}D^1(z).
\eqnum{22}
\end{equation}
We stress that these functions are singular solutions of the wave
equation in $3+1$ space, if $z^\alpha$ is construed as a
four-dimensional, rather than a two-dimensional, vector.  Here
the appearance of these functions, which depend on the 2-vector
$z^\alpha$, is a result of the deep symmetry between the creation of
a pair by a mirror in $1+1$ space and the emission of single
quanta by a charge in $3+1$ space.

Using
\begin{eqnarray}
du\,dv&=&d \tau\,d \tau '\dot{x}_-\dot{x}'_+=
d \tau\,d \tau '\left[
\frac{1}{2}\left( \dot{x}_-\dot{x}'_++\dot{x}_+\dot{x}'_-
\right)+
\frac{1}{2}\left( \dot{x}_-\dot{x}'_+-\dot{x}_+\dot{x}'_- \right)
  \right]=\nonumber\\
&=&
d \tau\,d \tau '
\left(
-\dot{x}_\alpha\dot{x}'{}^\alpha
+\varepsilon_{\alpha\beta}\dot{x}^\alpha
\dot{x}'{}^\beta \right)  \eqnum{23}
\end{eqnarray}
in the form of a sum of terms which are even and odd with respect
to the interchange $\tau \rightleftarrows \tau '$ (a dot denotes
differentiation with respect to the proper time), we obtain
\begin{equation}
N^B=\mathop{\int\!\!\!\int} \limits_{-\infty\quad}^{\quad\infty}
d\tau\,d\tau '
\left( \dot{x}_\alpha\dot{x}'{}^\alpha-\varepsilon_{\alpha\beta}
\dot{x}^\alpha\dot{x}'{}^\beta \right)
\frac{1}{2}D^1(z)\bigg|^F_0.     \eqnum{24}
\end{equation}

It is natural to use an explicitly relativistic method that
conserves the symmetry relative to the interchange $\tau
\rightleftarrows \tau '$ to eliminate the infrared divergence in
(24).  It consists of replacing the function $D^1\left( z \right)$ by
the function $\Delta^1(z,\mu)$, which is also even in $z$ and
has the small mass parameter $\mu\ll\kappa$, where $\kappa$ is the
characteristic acceleration of the mirror.

This function
\begin{equation}
\frac{1}{2}\Delta^1(z,\mu)=\frac{\mu}{8\pi s}N_1(\mu s)=
-\frac{1}{4\pi^2s^2}-\frac{\mu}{4\pi^2s}J_1(\mu s)
\ln\frac{2}{\mu s}+R      \eqnum{25}
\end{equation}
(where $J_1$ and $N_1$ are Bessel and Neumann functions, and $R$
is a regular function of $s$) is a singular solution of the wave
equation in $3+1$ space, which depends only on the interval
$s=\sqrt{-z^2}$ between the two points and preserves all the
features with respect to $s$ at $s=0$.  It is called the Hadamard
elementary function or the fundamental solution\cite{5}.  The
coefficient in front of the logarithm, which is called the Riemann
function\cite{5},
is a regular function of $s$, which satisfies the same
equation as $\Delta^1$.  Just these two functions will define the
imaginary and real parts of the action change.

Thus,
\begin{equation}
N^B=\mathop{\int\!\!\!\int} \limits_{-\infty\quad}^{\quad\infty}
d\tau\,d\tau '
\left( \dot{x}_\alpha\dot{x}'{}^\alpha-\varepsilon_{\alpha\beta}
\dot{x}^\alpha\dot{x}'{}^\beta \right)
\frac{1}{2}\Delta^1(z,\mu)\bigg|^F_0.        \eqnum{26}
\end{equation}

In the expressions for $N^B$ the odd term is unessential, since
$D^1(z)$ and $\Delta^1(z,\mu)$ are even relative to the replacement $z
\rightarrow -z$.

Now regarding $N$ as the imaginary part of the doubled
action,
it is naturally to consider the function $(1/2)\Delta^1(z,\mu)$ as
imaginary
part of some function $F(z^2)$, which is taken on the real $z^2$
axis,
and which is analytic in the $z^2$ complex plane with a cut along
the $z^2\leq 0$
semiaxis, where Lorentz invariance allows it to still depend on
the sign
of $z^0$, and coincides with $(i/2)\Delta^1(z,\mu)$ at $z^2>0$.  Then
the transition
from $iN^B$ to $2W$ is equivalent to the analytic continuation
of $F(z^2)$
onto the real semiaxis $z^2\leq 0$.  It is well known\cite{6}
that the boundary value of such a function, which does not depend
on the sign of $z^0$ and is, therefore, even, is the limit from above
($\varepsilon \rightarrow +0$), which is called a causal function:
\begin{equation}
\Delta_f(z,\mu)=F(z^2+i\varepsilon)=\frac{\mu}{4\pi^2s}K_1(i\mu s)=
\frac{1}{4\pi}\delta(s^2)-\frac{\mu}{8\pi s}
\left[J_1(\mu s)-iN_1(\mu s)  \right].  \eqnum{27}
\end{equation}
Here $K_1$ is the McDonald function, and
$s=\sqrt{-z^2-i\varepsilon}$.  The latter equality was written
for $z^2\leq 0$, where $s\geq 0$ and $\Delta_f$ has a real part, which
coincides with the Riemann function multiplied by $\pi/2$.  If
$z^2>0$, then $s=-i\sqrt{z^2}$, $\Delta_f$ is purely imaginary, and
its imaginary part is positive.

Thus, for $\Delta W_f^B$ we obtain
\begin{equation}
\Delta W^B_f=\frac{1}{2}
\int\!\!\!\int d \tau\,d \tau '\dot{x}_\alpha(\tau)
\dot{x}^\alpha(\tau ')\Delta_f(z,\mu)\bigg|^F_0.  \eqnum{28}
\end{equation}

As was shown in Ref. \onlinecite{2}, the space-time
representation for $N^F$ differs from the representation (17) for
$N^B$ by the additional multiplier $-\sqrt{f'(u)g'(v)}$ under the
integral.  Therefore, after the replacement of variables (18),
instead of (23) we have
\begin{equation}
-du\,dv\sqrt{f'(u)g'(v)}\,=-d \tau\,d \tau '.  \eqnum{29}
\end{equation}
Then
\begin{equation}
N^F=\frac{1}{2}
\int\!\!\!\int d\tau\,d\tau '
\Delta^1(z,\mu)\bigg|^F_0,   \eqnum{30}
\end{equation}
and the action change equals
\begin{equation}
\Delta W^F_f=\frac{1}{2}
\int\!\!\!\int d\tau\,d\tau '
\Delta_f(z,\mu)\bigg|^F_0.   \eqnum{31}
\end{equation}

The proper-time representations obtained for $\Delta
W_f^{B,F}$
differ from the self-action changes $\Delta W_1$ and $\Delta W_0$
of electric and scalar charges moving along the same
trajectories
as the mirror, but in $3+1$ space, only by the absence of the
multiplier $e^2$.

At $\mu \rightarrow 0$, the coefficients in front of $\ln
\mu^{-1}$ in the imaginary parts of the proper-time integrals
(28) and (31) should coincide with the coefficients in front of
$\ln L$ in the corresponding expressions for $N^B$ and $N^F$ [see
(15) and (16)], since these coefficients cannot depend on the
method used to eliminate the infrared divergence in the different
representations for each of quantities $N^B$ and $N^F$.

Since for an interval between two points on the timelike
trajectory the
\begin{equation}
\mathop{\rm Re}\Delta_f(z,\mu)=
-\frac{\mu}{8\pi s}J_1(\mu s)  \eqnum{32}
\end{equation}
and differs from the coefficient in front of the logarithm in
$\mathop{\rm Im}\Delta_f$ only by the multiplier $\pi/2$ [see (27) and
(25)], $\mathop{\rm Re}\Delta W_f$ also differs by the same multiplier
from the coefficient in front of $\ln\mu^{-1}$ in $\mathop{\rm
Im}\Delta W_f$.  Thus, to within terms which vanish at $\mu
\rightarrow 0$, we have
\begin{equation}
\Delta W_f=\pi a(\theta)+i\left[a(\theta)\ln\frac{\kappa^2}{\mu^2}+
b(\theta)  \right],     \eqnum{33}
\end{equation}
\begin{equation}
a^B(\theta)
=\frac{1}{8\pi^2}\left( \frac{\theta}{\tanh \theta}-1 \right),\quad
a^F(\theta)=\frac{1}{8\pi^2}\left( 1-\frac{\theta}{\sinh \theta}
\right).\eqnum{34}
\end{equation}
The function $b(\theta)$ can depend on other dimensionless parameters,
for example, the velocity changes on parts of the trajectory
containing other extreme values of the proper acceleration.

It is significant
that for $\theta\neq 0$ the $\mathop{\rm Re}\Delta W_f=\pi a$
has a finite positive limit at $\mu \rightarrow 0$.

To conclude this section we recall that $\mathop{\rm Re}\Delta
W_f$ is the acceleration-induced self-energy shift of the source
integrated over the proper time and that $2\mathop{\rm Im}\Delta
W_f$ is the mean number of emitted pairs (or emitted particles in
the case of their nonidentity to the antiparticles).  More precisely,
$\exp(-2\mathop{\rm Im}\Delta W_f)$ is the probability of the
noncreation of pairs during the all time of acceleration.

\section{ACTION
CHANGE IN THE CASE OF QUASIHYPERBOLIC MOTION OF MIRROR}

It would be interesting to directly calculate $\Delta W_f^{B,F}$
for the special, but very important mirror trajectory
\begin{equation}
x=\xi(t)=v_\infty\sqrt{\frac{v^2_\infty}{\kappa^2}+t^2},
\eqnum{35}
\end{equation}
which can be called quasihyperbolic.  Here $\pm v_\infty$ are the
velocities of the mirror at $t \rightarrow \pm\infty$, and
$\kappa$ is its acceleration at the turning point ($t=0$).  This
motion is remarkable in that as $v_\infty \rightarrow 1$, it
becomes increasingly close to uniformly accelerated (hyperbolic)
motion over the increasingly longer time interval
\[
|t|\lesssim t_1=\frac{v_\infty}{\kappa}(1-v^2_\infty)^{-1/2},
\]
smoothly going over to uniform motion outside this interval.
This can be seen from the expression for the magnitude of the
acceleration in the proper frame
\[
a=\kappa\left( 1+\frac{t^2}{t^2_1} \right)^{-3/2}.
\]
The spectrum and total radiated energy were found for an electric
charge moving along the trajectory (35) in Ref. \onlinecite{7}.

To calculate $\Delta W^B$, in (28) instead of $t$ we use the
variable $u$, which is defined by the formula
\[
 t=\frac{v_\infty}{\kappa}\sinh u.
\]

Then
\[
d \tau\,d \tau '\dot{x}_\alpha(\tau)\dot{x}^\alpha(\tau ')=-dx\,dy
 \frac{v^2_\infty}{\kappa^2}
 \left(
\frac{1+v^2_\infty}{2}\cosh x+\frac{1-v^2_\infty}{2}\cosh 2y
   \right),
\]
\begin{equation}
\eqnum{36}
\end{equation}
\[
(x-x')^2=-2\frac{v^2_\infty}{\kappa^2}(\cosh x-1)
 \left[v^2_\infty+(1-v^2_\infty)\cosh^2y  \right],\quad
 x=u-u',\quad
 y=\frac{u+u'}{2},
\]
and $\Delta W^B$ is expressed by the integral of the McDonald function
\begin{equation}
 \Delta W^B={-}\frac{1}{32\pi^2}
 \int\limits_{0}^{\infty}\frac{d\xi}{\xi^2}e^{{-}i\mu^2\xi}
 \int\limits_{{-}\infty}^{\infty}dy\,
 \frac{v^2_\infty}{\kappa^2}e^{iz}
 \left[(1 + v^2_\infty)K_1(iz) + (1 - v^2_\infty)K_0(iz)\cosh 2y
\right]^F_0,\eqnum{37}
\end{equation}
if we use the representation
\begin{equation}
 \Delta_f(x-x',\mu)=\frac{1}{16\pi^2}
 \int\limits_{0}^{\infty}\frac{d\xi}{\xi^2}\exp \left[
 i\frac{(x-x')^2}{4\xi}-i\mu^2\xi
    \right]     \eqnum{38}
\end{equation}
for the causal function and introduce the notation
\begin{equation}
z=\frac{v^2_\infty}{2\kappa^2\xi}
\left[v^2_\infty+(1-v^2_\infty)\cosh^2y  \right]. \eqnum{39}
\end{equation}
Now going over from the integration variable $\xi$ to $z$ in (37),
we obtain
\begin{equation}
\Delta W^B=-\frac{1}{8\pi^2}\int\limits_{-\infty}^{\infty}dy\left\{
\frac{1+v^2_\infty}{2Q}\left[S_1(\Lambda)+S_0(\Lambda)  \right]
-S_0(\Lambda)
  \right\},     \eqnum{40}
\end{equation}
where
\begin{equation}
\Lambda=\lambda v^2_\infty Q,\quad
\lambda=\frac{\mu^2}{\kappa^2},\quad
Q=v^2_\infty+(1-v^2_\infty)\cosh^2y,      \eqnum{41}
\end{equation}
\begin{equation}
S_n(\Lambda)=(-1)^{n+1}\int\limits_{0}^{\infty}dz\,e^{-i\Lambda/2z}
\left[
e^{iz}K_n(iz)-\sqrt{\frac{\pi}{2iz}}
  \right].      \eqnum{42}
\end{equation}

The subtraction $\big|_0^F$ in (40) was reduced to
subtraction of the asymptotics $(\pi/2iz)^{1/2}$ of the integrand in
(42) for $S_n(\Lambda)$.  As was shown in Refs. \onlinecite{8} and
\onlinecite{9}, the functions $S_n(\Lambda)$ are expressed in terms of
the product of the modified Bessel functions $I_n(\sqrt{\Lambda})$ and
$K_n(\sqrt{\Lambda})$.  We also turn attention to the more compact
expressions for the derivatives
\begin{equation}
S'_n(\Lambda)=(-1)^n\pi\left[
I_n(x)K_n(x)-\frac{1}{2x}
  \right]+iK^2_n(x),\quad
x=\sqrt{\Lambda}.     \eqnum{43}
\end{equation}
It can be seen from formulas (40)--(42) that $\Delta W^B$ depends on
two dimensionless parameters, viz., $\lambda$ and
$v_\infty=\tanh(\theta/2)$.

To calculate the asymptotics of the integral (40) at $\lambda
\rightarrow 0$
we note that in this case the values $\Lambda \rightarrow 0$
will be effective in the first term and, therefore,
\begin{equation}
S_1(\Lambda)+
S_0(\Lambda)\approx -\pi-i\ln\frac{4}{\gamma^2\Lambda},\quad
\gamma=1.781\ldots, \eqnum{44}
\end{equation}
and that in the second term the integral can be reduced to the
expression
\begin{eqnarray}
\int\limits_{-\infty}^{\infty}dy\,S_0(\Lambda)&\approx &
-\int\limits_{0}^{\infty}d \Lambda\,S'_0(\Lambda)\ln \Lambda+
S_0(0)\ln\frac{4}{\lambda v^2_\infty(1-v^2_\infty)}=
            \nonumber\\
&=&
-\pi-i\left[\ln\frac{16}{\gamma^2v^2_\infty(1-v^2_\infty)\lambda}-2
\right].
                    \eqnum{45}
\end{eqnarray}
As a result, to
within terms which vanish at $\lambda \rightarrow 0$ we
obtain
\begin{eqnarray}
\Delta W^B&\approx& \frac{1}{8\pi^2}\left\{
\pi\left( \frac{\theta}{\tanh \theta}-1 \right)+i\left[
\left( \frac{\theta}{\tanh \theta}-1 \right)
\ln\left[\frac{8(
\cosh \theta+1)^2}{\gamma^2\lambda(\cosh \theta-1)}\right]
+\right.\right.\nonumber\\
&+&2-
\left.
\left.
\frac{L_2(1-e^{-2\theta})+\theta^2}{\tanh \theta}
  \right]
  \right\}.     \eqnum{46}
\end{eqnarray}
Here $\theta=\tanh^{-1}
\beta_{21}=2\tanh^{-1} v_\infty$, and $L_2(x)$ is a Euler
dilogarithm.\cite{10,11}

For a quasiuniformly accelerated mirror interacting with a
spinor field, instead of (40) we obtain
\begin{equation}
\Delta W^F=\frac{1}{8\pi^2}\int\limits_{-\infty}^{\infty}dy
\left\{
\int\limits_{0}^{\infty}
dz\,\exp \left( -\frac{i\Lambda}{2z}+iz \right)
R(iz)-S_0(\Lambda)
  \right\},     \eqnum{47}
\end{equation}
where
\[
R(iz)=\int\limits_{0}^{\infty}dx\left(
\sqrt{\frac{(\cosh x+c)^2-s^2}{(1+c)^2-s^2}}-1
 \right)
\exp \left( -iz\cosh x \right),
\]
\begin{equation}
\eqnum{48}
\end{equation}
\[
c=\cosh \theta\cosh 2y,\quad
s=\sinh \theta\sinh 2y,\quad
\theta=2\tanh^{-1} v_\infty,
\]
and the remaining notation is the same as in (40).  It is seen
that
$\Delta W^F$ depends on the two dimensionless parameters $\lambda$ and
$\theta$.

When $\lambda \rightarrow 0$, the expression in the large
parentheses in (48) can be replaced by
\[
\frac{\cosh x-1}{\sqrt{(1+c)^2-s^2}}.
\]
This approximation holds for $\cosh x\gg 1$ and has the correct
(zero) value at $x = 0$. Then
\begin{equation}
\Delta W^F\approx \frac{1}{8\pi^2}
\int\limits_{-\infty}^{\infty}dy\left\{
\frac{1}{\sqrt{(1+c)^2-s^2}}
\left[S_1(\Lambda)+S_0(\Lambda)  \right]-S_0(\Lambda)
  \right\},      \eqnum{49}
\end{equation}
and using (44) and (45), we obtain
\begin{equation}
\Delta W^F\approx \frac{1}{8\pi^2}\left\{
\pi\left( 1{-}\frac{\theta}{\sinh \theta}\right){+}i\left[
\left( 1{-}\frac{\theta}{\sinh \theta} \right)
\ln\left[\frac{8
(\cosh \theta{+}1)^2}{\gamma^2\lambda(\cosh \theta{-}1)}\right]{-}2{+}
\frac{L_2(1{-}e^{{-}2\theta}){+}\theta^2}{\sinh \theta}
  \right]
  \right\}      \eqnum{50}
\end{equation}
to within terms which vanish at $\lambda \rightarrow 0$.

The formulas obtained for $\Delta W^{B,F}$ not only have the
structure (33), but also contain explicit expressions for
$b^{B,F}(\theta)$.
It can also be seen that $\Delta W^{B,F}$ do not depend
on the sign of $\theta$ or $\beta_{21}$, if we take into account that
$L_2(1-e^{-2\theta})+\theta^2$
is an odd function of $\theta$ [see Landen's
formula (1.12) in Ref. \onlinecite{11}].  We note in this
connection that for small values of $\theta$
\begin{equation}
L_2(1-e^{-2\theta})+\theta^2=2\theta+\frac{2}{9}\theta^3-
\frac{2}{225}\theta^5+\ldots,       \eqnum{51}
\end{equation}
and that as $\theta \rightarrow \pm\infty$, to within exponentially
small terms we have
\begin{equation}
L_2(1-e^{-2\theta})+\theta^2=\pm\left(
\theta^2+\frac{\pi^2}{6}\right)+\ldots.     \eqnum{52}
\end{equation}

The imaginary and real parts of $\Delta W^{B,F}$ in (46) and (50)
are positive owing to unitarity and causality.  When
$\theta=0$, $\Delta W$ vanishes, since the quasihyperbolic trajectory
becomes straight line.

The point
$\theta=\infty$ for $\Delta W^{B,F}_f(\theta,\lambda)$ is essential
singularity.  It physically corresponds to a purely hyperbolic
trajectory for which $\beta_{21}=1$ or $- 1$ in accordance with the
sign of $\kappa$.
At a fixed value of $\lambda$ and $\theta \rightarrow
\pm\infty$, from (40) and (47) we obtain
\begin{equation}
\Delta W^{B,F}_f(\theta,\lambda)=
\mp \theta\frac{1}{8\pi^2}S_{1,0}(\lambda).
\eqnum{53}
\end{equation}
Here $\pm \theta=|\kappa|(\tau_2-\tau_1)\gg 1$, and when the length of
proper time interval $(\tau_1,\tau_2)$ approaches to infinity the
relative
velocity $\beta_{21}$ approaches $+1$ or $-1$.  Formula (53) was
obtained for uniformly accelerated charges in $3+1$ space in
Ref.  \onlinecite{12}
and was discussed in detail in Refs.  \onlinecite{8}
and \onlinecite{9}.
In those studies it defined the classical mass shift
of a uniformly accelerated charge:
\begin{equation}
\Delta m_{1,0}=-\frac{\partial \Delta W_{1,0}}{\partial
\tau_2}= \frac{\alpha}{2\pi}|\kappa|S_{1,0}(\lambda).  \eqnum{54}
\end{equation}

In accordance
with the unitarity and causality, the imaginary and real parts
of $\Delta m$
are negative.  At $\kappa=0$ the function $\Delta m(\kappa)$
is nonanalytic
and, therefore, cannot be reproduced by perturbation theory
with respect to $\kappa$ or with respect to the field accelerating the
charge.

\section{DISPERSION RELATIONS FOR $\Delta W$ AND THEIR ORIGIN}

It was shown in Ref. \onlinecite{9} that the action changes
$\Delta W_s(\mu^2)$ of point charges moving along timelike
trajectories as functions of the square of the mass of quanta of
their proper-field with spin $s = 1,0$ are analytic in the $\mu^2$
complex plane with a cut along the positive $\mu^2$ semiaxis,
on whose edges the imaginary parts of each of the functions
coincide, while the real parts differ in sign.  Such functions
satisfy the dispersion representations ($\mathop{\rm Im}\mu<0$ ):
\begin{equation}
\Delta W(\mu^2)=\frac{2i}{\pi}
\int\limits_{0}^{\infty}\frac{dx\,x\mathop{\rm Re}\Delta
W(x^2)}{x^2-\mu^2}=-
\frac{2\mu}{\pi}
\int\limits_{0}^{\infty}\frac{dx\mathop{\rm Im}\Delta
W(x^2)}{x^2-\mu^2},   \eqnum{55}
\end{equation}
which reconstruct the function $\Delta W_s(\mu^2)$ in the
$\mu^2$ complex plane from its real or imaginary part assigned on
the lower edge of the cut.  When $\mu=i\kappa$ and
$\kappa>0$, these relations yield the important equalities
\begin{equation}
\frac{2}{\pi}\int\limits_{0}^{\infty}
\frac{dx\,x\mathop{\rm Re}\Delta W(x^2)}{x^2+\kappa^2}=
\frac{2\kappa}{\pi}
\int\limits_{0}^{\infty}\frac{dx\mathop{\rm Im}\Delta
W(x^2)}{x^2+\kappa^2}=
\mathop{\rm Im}\Delta W(-\kappa^2)>0,       \eqnum{56}
\end{equation}
\begin{equation}
\mathop{\rm Re}\Delta W(-\kappa^2)=0.       \eqnum{57}
\end{equation}

As a consequence of unitarity, the $\mathop{\rm Im}\Delta
W(\mu^2)$ is positive on the real semiaxis $\mu^2>0$.  Then,
according to the second of the representations (55), $\mathop{\rm
Im}\Delta W(\mu^2)$ is positive definite over the entire $\mu^2$
complex plane (or in the lower $\mu$ half-plane).

Here we show that the dispersion relations presented for $\Delta
W(\mu^2)$ are due to the analytic properties of the causal
Green's function $\Delta_f(z,\mu)$, which, as we see, specifies not
only $\Delta W_s(\mu^2)$ for the vacuum amplitude of accelerated
charges in $3+1$ space, but also $\Delta W^{B,F}(\mu^2)$ for the
vacuum amplitude of an accelerated mirror in $1+1$ space.

We can show that the causal function $\Delta_f(z,\mu)$ for a
timelike $z$ satisfies the dispersion relations presented.
According to formulas (2.12.4.28) and (2.13.3.20) from Ref.
\onlinecite{13}
\begin{equation}
\int\limits_{0}^{\infty}\frac{dx\,x^2J_1(sx)}{x^2+\kappa^2}=
-\kappa
\int\limits_{0}^{\infty}\frac{dx\,xN_1(sx)}{x^2+\kappa^2}=
\kappa K_1(s\kappa),        \eqnum{58}
\end{equation}
where $s, \mathop{\rm Re}\kappa>0$.  After the analytic
continuation
in $\kappa$ to the point $\kappa=i\mu+\varepsilon$ where $\mu>0$
and $\varepsilon \rightarrow +0$ these relations turn into
\begin{equation}
\int\limits_{0}^{\infty}\frac{dx\,x^2J_1(sx)}{x^2-\mu^2
+i\varepsilon}=-i\mu
\int\limits_{0}^{\infty}\frac{dx\,xN_1(sx)}{x^2
-\mu^2+i\varepsilon}=i\mu
K_1(i\mu s)=
-\frac{i\pi\mu}{2}
\left[J_1(\mu s)-iN_1(\mu s)  \right].  \eqnum{59}
\end{equation}
After multiplying by $-i/4\pi^2s$, they form the first pair of
dispersion relations (55), which, instead of $\Delta W(\mu^2)$,
contain the causal function (27) with a timelike vector $z^\alpha$,
for which $s=\sqrt{-z^2}>0$.  For spacelike $z^\alpha$ the interval
$s=-i\sqrt{z^2}$, and $\Delta_f(z,\mu)$ is purely imaginary.

After being multiplied by $-1/4\pi^2s$, the original formulas
(58) coincide with the second pair of the relations (56) with the
replacement of $\Delta W(\mu^2)$ by $\Delta_f(z,\mu)$.  The function
appearing on the right-hand side of these relations
\begin{equation}
-\frac{\kappa}{4\pi^2s}K_1(\kappa s) = \mathop{\rm
Im}\Delta_f(z,-i\kappa), \eqnum{60}
\end{equation}
unlike $\mathop{\rm Im}\Delta W(-\kappa^2)$, is negative.  In
addition,
\begin{equation}
\mathop{\rm Re}\Delta_f(z,-i\kappa)=0,  \eqnum{61}
\end{equation}
as can be seen from (27).  This property is a consequence of the
causality, according to which $\mathop{\rm Re}\Delta_f(z,\mu)=0$
outside the light cone, i.e., for spacelike $z^\alpha$.  In this case
the argument of the McDonald function in (27) is real and
positive.  When we go over to timelike $z^\alpha$ and a purely
imaginary negative $\mu=-i\kappa$, this argument remains real and
positive, whence follows (61).

While satisfying the dispersion relations (55) and (56) with
respect to the ``dispersion'' variable $\mu$, the function
$\Delta_f(z,\mu)$,
unlike $\Delta W(\mu^2)$, still depends on the fixed
as yet parameter $s$, which equals the invariant
interval between the two points chosen on the mirror trajectory
with the
proper times $\tau$ and $\tau '$, i.e., on $s=s(\tau,\tau ')$.
Integrating
the dispersion relations for $\Delta_f$ over $\tau, \tau '$ with
the weight
$(1/2)\dot{x}_\alpha(\tau)\dot{x}^\alpha(\tau ')$ or $1/2$ and
performing the subtraction procedure, we obtain the dispersion
relations
for $\Delta W^B$ or $\Delta W^F$, if, of course, the familiar
conditions for changing the order of integration over $x$ and
$\tau,\tau '$ are satisfied.

Thus, the dispersion relations for $\Delta W(\mu^2)$ are a
consequence of the dispersion relations for $\Delta_f(z,\mu)$.

If follows from (56) that if $\mathop{\rm Im}\Delta W(\mu^2)$ is
bound at zero, then $\mathop{\rm Re}\Delta W(\mu^2)$ must vanish at
$\mu \rightarrow +0$.  If, on the other hand, at $\mu \rightarrow 0$
the $\mathop{\rm Im}\Delta W(\mu^2)$ logarithmically tends to infinity
according to the relation
\begin{equation}
\mathop{\rm Im}\Delta W(\mu^2)=a\ln\mu^{-2}+b(\mu^2)    \eqnum{62}
\end{equation}
[$a>0$, and $b(\mu^2)$ is bound at zero], it follows from (56) that
$\mathop{\rm Re}\Delta W(\mu^2)$ tends to the positive value
$\mathop{\rm Re}\Delta W(0)=\pi a$ at $\mu \rightarrow 0$.
According to (57),
this means that the function $\mathop{\rm Re}\Delta W(\mu^2)$ has a
discontinuity equal to $\pi a$ on the real $\mu^2$ axis at $\mu^2 =0$.

\section{INFLUENCE OF THE BOUNDARY CONDITIONS ON ${\mathop{\rm
R\lowercase{e}}\Delta W}$}

Let us now consider the other boundary values of $F(z^2)$,
which is analytic in the $z^2$ complex plane with a cut along
the $z^2\leq 0$ semiaxis and coincides with $(i/2)\Delta^1(z,\mu)$ on
the $z^2>0$ semiaxis.

The limit $F(z^2-i\varepsilon)$ on the real axis from below is
distinguished from the limit (27) from above by the opposite sign of
the real part.  According to this function, free fields would
transfere negative energy in $3+1$ space; therefore, this
boundary condition is  not considered here.

The other boundary values of $F(z^2)$, which already depend
on the sign of $z^0$, may be the limits $F(z^2 \pm
i\varepsilon\mathop{\rm sgn} z^0)$, $\varepsilon \rightarrow +0$.
They are positive- and negative-frequency functions, or, more
precisely, $\pm\Delta^\pm(z,\mu)$ (Ref. \onlinecite{4}):
\begin{equation}
\pm \Delta^\pm(z,\mu)=\pm\varepsilon(z^0)\mathop{\rm
Re}\Delta_f+i\mathop{\rm Im}\Delta_f.
\eqnum{63}
\end{equation}
Such functions naturally vary only the real part of the action
obtained for $\Delta_f$, so that
\begin{equation}
\mathop{\rm Re}\Delta W^B_\pm=\pm\frac{1}{2}
\int\!\!\!\int
d \tau\,d \tau '\left(
\dot{x}_\alpha\dot{x}'{}^\alpha
-\varepsilon_{\alpha\beta}\dot{x}^\alpha
\dot{x}'{}^\beta
\right)
\mathop{\rm Re}\Delta^\pm(z,\mu)\bigg|^F_0       \eqnum{64}
\end{equation}
differ from $\mathop{\rm Re}\Delta W^B_f$ and are given in the limit
$\mu \rightarrow 0$ by the expressions
\begin{equation}
\mathop{\rm Re}\Delta W^B_\pm=\mp\frac{1}{8\pi}
\int\!\!\!\int
d \tau\,d \tau '
\varepsilon_{\alpha\beta}
\dot{x}^\alpha\dot{x}'{}^\beta
\varepsilon(z^0)\delta(z^2).
       \eqnum{65}
\end{equation}
The integrand
can be expanded in $\tau '$ near $\tau '=\tau$ and represented
in the form
\begin{equation}
\varepsilon_{
\alpha\beta}\dot{x}^\alpha\dot{x}'{}^\beta\varepsilon(z^0)\delta(z^2)=
-\varepsilon_{
\alpha\beta}\dot{x}^\alpha\ddot{x}^\beta \delta(\tau-\tau ').
\eqnum{66}
\end{equation}
Here
the equality $|x|\delta(x^2)=\delta(x)$ was used (see, for example,
Ref. \onlinecite{14}).

Then, integrating over $\tau '$ and expressing the
proper-acceleration
\begin{equation}
a(\tau)=\varepsilon_{\alpha\beta}\dot{x}^\alpha\ddot{x}^\beta=
\frac{f''}{2(f')^{3/2}}=
\frac{d\ln f'(u)}{2d\tau}=
\frac{d\tanh^{-1} \beta(\tau)}{d \tau}       \eqnum{67}
\end{equation}
in the form of the derivative of the rapidity with respect to the
proper time, we obtain
\begin{equation}
\mathop{\rm Re}\Delta W^B_\pm=\pm\frac{1}{8\pi}
\int\limits_{-\infty}^{\infty}
d \tau\,\varepsilon_{\alpha\beta}\dot{x}^\alpha\ddot{x}^\beta=\pm
\frac{1}{8\pi}\tanh^{-1} \beta_{21}=\pm\frac{\theta}{8\pi}. \eqnum{68}
\end{equation}
Clearly,
\begin{equation}
\mathop{\rm Re}\Delta W^F_\pm=\pm\frac{1}{2}\int\!\!\!\int
d \tau\,d \tau '\mathop{\rm Re}\Delta^\pm(z,\mu)=0     \eqnum{69}
\end{equation}
because of the oddness of $\mathop{\rm Re}\Delta^\pm$ with respect to
$z$.

The expressions obtained for $\mathop{\rm Re}\Delta W_\mp$
up to the multiplier $(8\pi)^{-1}$ coincide with the odd in $\theta$
coefficients for the terms
proportional to $u^{-1}$ and $v^{-1}$ in the asymptotic
expansions of $K(u)$ and $K(v)$, respectively [see (13) and (14)
and the comment following Eq. (16)].  At the same time,
$\mathop{\rm Re}\Delta W_f$ up to the same multiplier
$(8\pi)^{-1}$ coincides with the even in $\theta$ coefficient
for the term proportional to $u^{-1}$ or $v^{-1}$ in the
asymptotic
expansion of $K(u)$ or $K(v)$.  We note that all these coefficients,
as well as
the functions $K(u)$ and $K(v)$ themselves, are formed without any
involvement
of the parameter $L$, which eliminates the infrared divergence of
the space-time integrals (9) for the mean number of particles emitted.

Thus, information on the interaction contained in $K(u)$ and
$K(v)$, which determine $\mathop{\rm Im}\Delta W$, is conveyed to
$\mathop{\rm Re}\Delta W$ owing to the causality and the boundary
conditions.  In addition, $\mathop{\rm Re}\Delta W_f$ contains
information on the interaction which propagates within the light
cone, and $\mathop{\rm Re}\Delta W_\pm$ contains information on the
interaction which propagates along the light cone and is
therefore local owing to the timelike character of the
trajectory.

As we know,\cite{4} the half-sum of the retarded and advanced
fields is the proper-field of the source, and their half-difference
is the radiation field escaping to infinity.  Since
\[
\mathop{\rm Re}\Delta_f
=\frac{1}{2}\left( \Delta^{\rm ret}+\Delta^{\rm adv}
\right),
\]
and
\[
\mathop{\rm Re}\Delta^+
=\frac{1}{2}\left( \Delta^{\rm ret}-\Delta^{\rm adv}
\right),
\]
$\mathop{\rm Re}\Delta W_f$ describes the self-energy shift of the
source, and $\mathop{\rm Re}\Delta W_+$ describes the interaction with
the radiation field, i.e., with real quanta.  The boundary
condition which eliminates the interaction with virtual quanta or
pairs seems unnatural.

\section{DISCUSSION AND PHYSICAL INTERPRETATION OF RESULTS}

The proper-time representations for the changes in the
self-interaction of a mirror upon acceleration in a
two-dimensional vacuum of scalar and spinor fields can be
considered the most significant results of this work.  These
representations coincided with the representations for the changes
in the self-interaction of electric and scalar charges
accelerated in four-dimensional space-time.  In other words, both
were found to be identical functionals of the source trajectory.

This coincidence, first, confirms the correctness of the
interpretation given in Ref. \onlinecite{3} of the Bogolyubov
coefficient
$\beta^*_{\omega '\omega}$ as the source amplitude of a virtual
pair of particles potentially emitted to the right and to the left
with the
frequencies $\omega$ and $\omega '$, with the timelike 2-momentum
of the pair (5), the mass $m=2\sqrt{\omega\omega'}$, and a spin equal
to 1 for a boson pair and 0 for a fermion pair.

Second, it means that the self-interaction of mirror
is realized by the creation and absorption of virtual pairs, rather
than individual particles, and is transfered from one point of
the trajectory to another by the causal Green's function of the
wave equation for four-dimensional, rather than two-dimensional
space-time.

The action integral is formed by virtual pairs with the mass
$m=2\sqrt{\omega\omega'}$, which takes any positive values.
Therefore, it is natural to expect that the effective propagation
function of such pairs will be the integral of the propagation
function of a massive particle in two-dimensional space-time over
the mass $m$.

At the same time, it can be shown that the causal Green's
functions for spaces of dimensionalities $d$ and $d+2$, being
functions of the invariant interval  $s=\sqrt{-z^2}$ between two
points and the mass $\mu$, are related to one another by the
equalities
\begin{equation}
\Delta_f^{(d+2)}(z,\mu)=\frac{1}{\pi}\frac{\partial }{\partial s^2}
\Delta_f^{(d)}(z,\mu)=\frac{1}{4\pi}
\int\limits_{\mu^2}^{\infty}
dm^2\,\Delta_f^{(d)}(z,m) \eqnum{70}
\end{equation}
and are expressed in terms of the McDonald function with the
index specified by the dimensionality of the space-time:
\begin{equation}
\Delta_f^{(d)}(z,\mu)=\frac{i\mu^{2\nu}}{(2\pi)^{\nu+1}(i\mu s)^\nu}
K_\nu(i\mu s),\quad
\nu=\frac{d-2}{2}.      \eqnum{71}
\end{equation}

The second equality in (70) for $d=2$ confirms the
appearance of a causal function characteristic of
four-dimensional space-time as an effective propagation function
of virtual pairs with different masses $m$ in two-dimensional
space-time.  Now the small mass parameter $\mu$, which was
introduced in Sec. 2 to eliminate the infrared divergence, can be
interpreted as the lower bound of the masses of the virtual pairs
transfering the self-interaction of a mirror.

A virtual pair can not escape to infinity, since one of its
particles unavoidably undergoes reflection from the mirror, after
which the pair becomes real and massless.  The emission of such
pairs forms $\mathop{\rm Im}\Delta W$.  Owing to masslessness,
the emission of arbitrary large number of arbitrary soft quanta
becomes possible on trajectories with $\beta_{21}\neq 0$,
i.e., infrared divergence of  $\mathop{\rm Im}\Delta W_f$ appears.  By
choosing a nonzero, but sufficiently small value for $\mu$, we
eliminate the infrared divergence  in $\mathop{\rm Im}\Delta W_f$ and
make sure that $\mathop{\rm Re}\Delta W_f$ does not depend on $\mu$ at
$\mu\ll|\kappa|$.  This means that the main contribution to
$\mathop{\rm Re}\Delta W_f$ is made by virtual pairs with a mass of
the order of $|\kappa|$.

In the general case, where the mean number of pairs created
is not small compared to 1, the quantity $2\mathop{\rm Im}\Delta W$ is
no longer equal to the mean number of pairs ${\rm tr} (\beta^+\beta)$.
Because
of the interference of two or more being created pairs it equals
\begin{equation}
2\mathop{\rm Im}\Delta W=\pm{\rm tr}\ln(1\pm\beta^+\beta)\big|^F_0=
\pm{\rm tr}\ln(\alpha^+\alpha)\big|^F_0.  \eqnum{72}
\end{equation}
The last formula prompted De~Witt\cite{15} to consider the following
expression for $W$ natural:
\begin{equation}
W=\pm i\mathop{\rm tr}\ln \alpha.     \eqnum{73}
\end{equation}
The matrix
formulation of the Bogolyubov coefficients $\alpha$ and $\beta$
was adopted.  In addition, ${\rm tr}$ must be replaced by
$(1/2){\rm tr}$ in the case of an identical particle and
antiparticle \cite{3}.

We do not know of any concrete results for $\mathop{\rm Re}\Delta
W$ emanating from (73).

The symmetry discussed would be total, if the equality
$e^2=\hbar c$ would hold in Heaviside units.

This work was carried out with the financial support of the
Russian Foundation for Basic Research (Grants Nos. 96-15-96463
and 99-02-17916).

\end{document}